\documentclass[preprint]{aastex62}  

\begin{document}

                           % The preamble begins here.
\title{Colors and Dynamics of a Near-Sun Orbital Asteroid Family: 2021 PH27 and 2025 GN1}
%\correspondingauthor{Scott S. Sheppard}
%\email{ssheppard@carnegiescience.edu}

\author[0000-0003-3145-8682]{Scott S. Sheppard} %Ready to Submit
\affil{Earth and Planets Laboratory, Carnegie Science, 5241 Broad Branch Rd. NW, Washington, DC 20015, USA, ssheppard@carnegiescience.edu}

\author[0000-0001-7225-9271]{Henry H. Hsieh} %Ready to Submit
\affil{Planetary Science Institute, 1700 East Fort Lowell Rd., Suite 106, Tucson, AZ 85719, USA}

\author[0000-0002-5667-9337]{Petr Pokorn\'{y}} %Ready to Submit
\affiliation{Department of Physics, The Catholic University of America, Washington, DC 20064, USA}
\affiliation{Astrophysics Science Division, NASA Goddard Space Flight Center, Greenbelt, MD 20771}
\affiliation{Center for Research and Exploration in Space Science and Technology, NASA/GSFC, Greenbelt, MD 20771}

\author[0000-0003-0773-1888]{David J. Tholen} %Ready to Submit
\affil{Institute for Astronomy, University of Hawai'i, Honolulu, HI 96822, USA}

\author[0000-0002-1506-4248]{Audrey Thirouin} %Ready to Submit
\affil{Lowell Observatory, 1400 W Mars Hill Road, Flagstaff, AZ 86001, USA}

\author{Carlos Contreras} %Ready to Submit
\affil{Las Campanas Observatory, Carnegie Science, Raul Bitran 1200, La Serena, Chile}

\author{Marcelo D. Mora} %Ready to Submit
\affil{Las Campanas Observatory, Carnegie Science, Raul Bitran 1200, La Serena, Chile}

\author{Mauricio Martinez} %Ready to Submit
\affil{Las Campanas Observatory, Carnegie Science, Raul Bitran 1200, La Serena, Chile}

\author{Ivonne Toro} %Ready to Submit
\affil{Las Campanas Observatory, Carnegie Science, Raul Bitran 1200, La Serena, Chile}

\begin{abstract}  % Produces abstract

We observed the dynamically similar near-Sun asteroids 2021 PH27 and
2025 GN1 for their optical colors. These objects have the lowest known
semi-major axes of any asteroids. 2021 PH27 has the largest general
relativistic effects of any known solar system object. The small
semi-major axis and very close passage to the Sun suggests the extreme
thermal and gravitational environment should highly modify these
asteroids' surfaces. From g', r', i' and z'-band imaging, we find the
colors of 2021 PH27 to be between the two major asteroid types the S
and C classes ($g'-r'= 0.58 \pm 0.02$, $r'-i'=0.12 \pm 0.02$ and
$i'-z'=-0.08 \pm 0.05$ mags). With a spectral slope of $6.8\pm 0.03$
percent per 100nm, 2021 PH27 is a X-type asteroid and requires albedo
or spectral features to further identify its composition. We find the
dynamically similar 2025 GN1 also has very similar colors ($g'-r'=0.55
\pm0.06$ and $r'-i'=0.14\pm 0.04$) as 2021 PH27, suggesting these
objects are fragments from a once larger parent asteroid or 2021 PH27
is shedding material. The colors are not blue like some other near-Sun
asteroids such as 3200 Phaethon that have been interpreted to be from
the loss of reddening substances from the extreme temperatures. There
is no evidence of activity or a large amplitude period for 2021 PH27,
whereas 2025 GN1 might have a more significant rotational light
curve. 2025 GN1 may have a very close encounter or hit Venus in about
2155 years and likely separated from 2021 PH27 in the last $\sim 10$
kyrs.

\end{abstract}

\section{Introduction}

The near-Sun object 2021 PH27 was found in a deep and wide asteroid
twilight survey using the Dark Energy Camera on the Blanco 4 meter
telescope (Sheppard et al. 2022).  2021 PH27 is an Atira asteroid that
has an orbit completely interior to Earth's and the lowest known
semi-major axis of any asteroid. The orbital parameters from the Minor
Planet Center are $a=0.4617$ au, $e=0.7116$, and $i=31.9435$ deg.  The
perihelion of 2021 PH27 is very close to the Sun and interior to
Mercury's orbit at 0.1332 au, while the aphelion is just beyond Venus'
orbit at 0.790 au, giving an orbital period of 0.31 years. With an
absolute H magnitude near 17.6 mag, 2021 PH27 is about 1 km in
diameter assuming a moderate albedo. This likely makes 2021 PH27 one
of the top five largest near-Sun asteroids ($q< 0.15$ au) with only
3200 Phaethon significantly larger at an absolute magnitude of 14.4
mags and diameter of $\sim 5-6$ km (Taylor et al. 2019; Yoshida et
al. 2023). (137924) 2000 BD19, (394130) 2006 HY51 and (465402) 2008
HW1 are the only other low perihelia asteroids with brighter absolute
magnitudes, though they are all likely around 1 km in size like 2021
PH27.  2021 PH27 has such a low semi-major axis and a perihelion even
lower than Mercury, that it experiences the largest known general
relativistic effects for an object in our Solar System (Sheppard et
al. 2022). It has strong interactions with Venus and will likely pass
within a Hill sphere of Venus in the next few hundred years (Sheppard
et al. 2022; Carbognani et al. 2022).

2025 GN1 was discovered in April 2025 with a 5 night observational arc
published on MPEC 2025-G124 (Perkins et al. 2025).  We noticed the
orbit of 2025 GN1 was very similar to that of 2021 PH27, having
$a=0.462$ au, $e=0.71$, and $i=32.8$ deg. This suggests 2021 PH27 and
2025 GN1 might make a dynamical family from the breakup of a once
larger parent asteroid, as found for other asteroid pairs and families
(Nesvorny et al. 2024). Because of the dynamical similarities of these
two near-Sun objects, we have observed both for their optical colors,
obtained additional astrometry to extend the observational arc of 2025
GN1 as well as run numerical dynamical simulations to assess the
likelihood of both objects being from a once larger parent asteroid.

\section{Observations}

The 6.5 meter Magellan-Baade telescope with IMACS was used on the
nights of UT 2022 April 8 and 9 to measure the g', r', i' and z' Sloan
colors of 2021 PH27. The same setup at Magellan-Baade with IMACS was
again used on UT 2025 April 14 to measure the g', r', and i' Sloan
colors of 2025 GN1.  The IMACS camera is a mosaic imager with eight
$2048 \times 4096$ CCDs and a pixel scale of 0.2 arcseconds (Dressler
et al. 2011). The asteroids were centered on chip 2, which is near the
center of the field-of-view, for the observations. The airmass range
was 2.27 to 1.70 on April 8 and 2.16 to 1.75 on April 9 in 2022 with
seeing between 0.5 and 0.6 arcseconds on the two nights. On April 14
in 2025 the airmass range was between 2.7 and 1.7 airmasses with
seeing between 0.7 and 1 arcsecond. The Sun was about -34 and -23 deg
in altitude at the start in 2022 and 2025, respectively and near -14
deg altitude at the end of the observations, allowing for dark skies
during the images. A medianed bias image was subtracted from each
asteroid image with medianed twilight sky flats used to correct each
images sensitivity. All nights were photometric and the standard star
PG1633 was used as well as the background field stars using Refcat2 to
calibrate the Sloan color photometry (Landolt 2009; Tonry et
al. 2018).

The geometric circumstances of the observations for each night are
shown in Table~\ref{tab:GeoPhase}. Both 2021 PH27 and 2025 GN1 had
very similar observational circumstances.  2021 PH27 was about 0.76 au
from the Sun and 2025 GN1 0.75 au, both near their aphelia. The
asteroids had similar apparent motions of $234-238$ arcseconds per
hour East for 2021 PH27 and 258 for 2025 GN1 and $11-16$ and 24
arcseconds per hour North, respectively during the observations. The
telescope was guided at half the non-sidereal rates of the asteroids
in order to have the field stars and asteroid trailed a similar
amount. This half trailed technique allows the field stars to have
their photometry obtained in the same manner as the asteroid, allowing
for more accurate absolute calibrations. Images were 60 seconds for
2021 PH27 and 120 seconds for 2025 GN1 and rotated through the r', g',
i', and z' band filters to prevent any rotational variations of the
asteroids from significantly affecting the colors. 11 r'-band, 9
g'-band, 9 i'-band and 7 z'-band images were obtained for the
photometry of 2021 PH27 and 4 r'-band, 4 g'-band and 4 i'-band images
were obtained for 2025 GN1.

\section{Photometry Results}

2025 GN1 is about 2.3 mags fainter than 2021 PH27 in apparent
magnitude and absolute magnitude, making it about a third the size of
2021 PH27 if they have similar albedos.  Assuming an albedo of 0.15,
2021 PH27 is about 1.2 km in diameter and 2021 GN1 is about 400 m in
diameter. We find the average colors of 2021 PH27 as $g'-r'= 0.59 \pm
0.02$, $r'-i'=0.12 \pm 0.02$ and $i'-z'=-0.08 \pm 0.05$ mags
(Table~\ref{tab:Colors}). Very similar average Sloan colors to 2021
PH27 were found for 2025 GN1 of $g'-r'= 0.55 \pm 0.06$ and $r'-i'=0.14
\pm 0.04$. We did not obtain z'-band images of 2025 GN1 because it was
deemed too faint to have a reasonable signal-to-noise in the z'-band
filter and to allow for additional g', r' and i'-band
observations. The optical colors of 2021 PH27 and 2025 GN1 are
indistinguishable from each other, both being within the uncertainty
of each other in their Sloan optical colors, with spectral slopes of
$6.85 \pm 0.03$ and $6.80 \pm0.07$ percent per 100nm, respectively.

The dynamical and compositional similarity of 2021 PH27 and 2025 GN1
suggests they are both fragments from a once larger parent asteroid
that has fragmented or is shedding material from the extreme stress
environment near the Sun.  Several asteroid pairs have been physically
and dynamically studied to help determine their formation (Pravec et
al. 2019).  Other asteroids with dynamically similar orbits to each
other have also been found to have similar colors, which is expected
if the fragments of a break-up are all made of the same material from
the parent asteroid (Moskovitz et al. 2019; Jewitt et al. 2023).
Possible causes for the fragmentation could be from thermal stress,
tidal disruption, rotational spin-up or an escaped satellite (Pravec
et al. 2010; Jewitt et al. 2013; Granvik et al. 2018; Nesvorny et
al. 2023; Granvik \& Walsh 2024; Jewitt et al. 2025; Jewitt
2025). Sheppard et al. (2022) found 2021 PH27 has very strong
interactions with Venus, passing within a Hill Sphere of Venus in the
near future.  This means the likely dust created by 2021 PH27 and its
retinue of smaller objects like 2025 GN1 at break-up may create an
annual meteor shower at Venus like 3200 Phaethon does for Earth
(Carbognani et al. 2022).  Interestingly, Phaethon has also been found
to have known smaller asteroids in very similar orbits to it: 1999 YC
and 2005 UD (Jewitt \& Hsieh 2006; Cukier \& Szalay 2023).  Thus 2021
PH27 with 2025 GN1 might be to Venus like Phaethon and its known
dynamical family members 2005 UD and 1999 YC are to Earth.

Sloan asteroid colors show a clear distinction between the main S-type
and C-type asteroids, though there is some overlap (Parker et
al. 2008).  The colors of 2021 PH27 and 2025 GN1 are in between the
two main asteroid classes of S-type and C-type and they have the
colors of an X-type asteroid (Sergeyev \& Carry 2021). X-type is the
indeterminate color of an asteroid that requires further information
to properly categorize the surface such as albedo or spectral
features. The X-type Sloan color is between the main C-type and S-type
asteroids and believed to contain E-type, M-type and P-type asteroids
(DeMeo \& Carry 2013).

\begin{figure}
  \epsscale{0.4}
  \centering
\includegraphics[angle=180,totalheight=0.3\textheight]{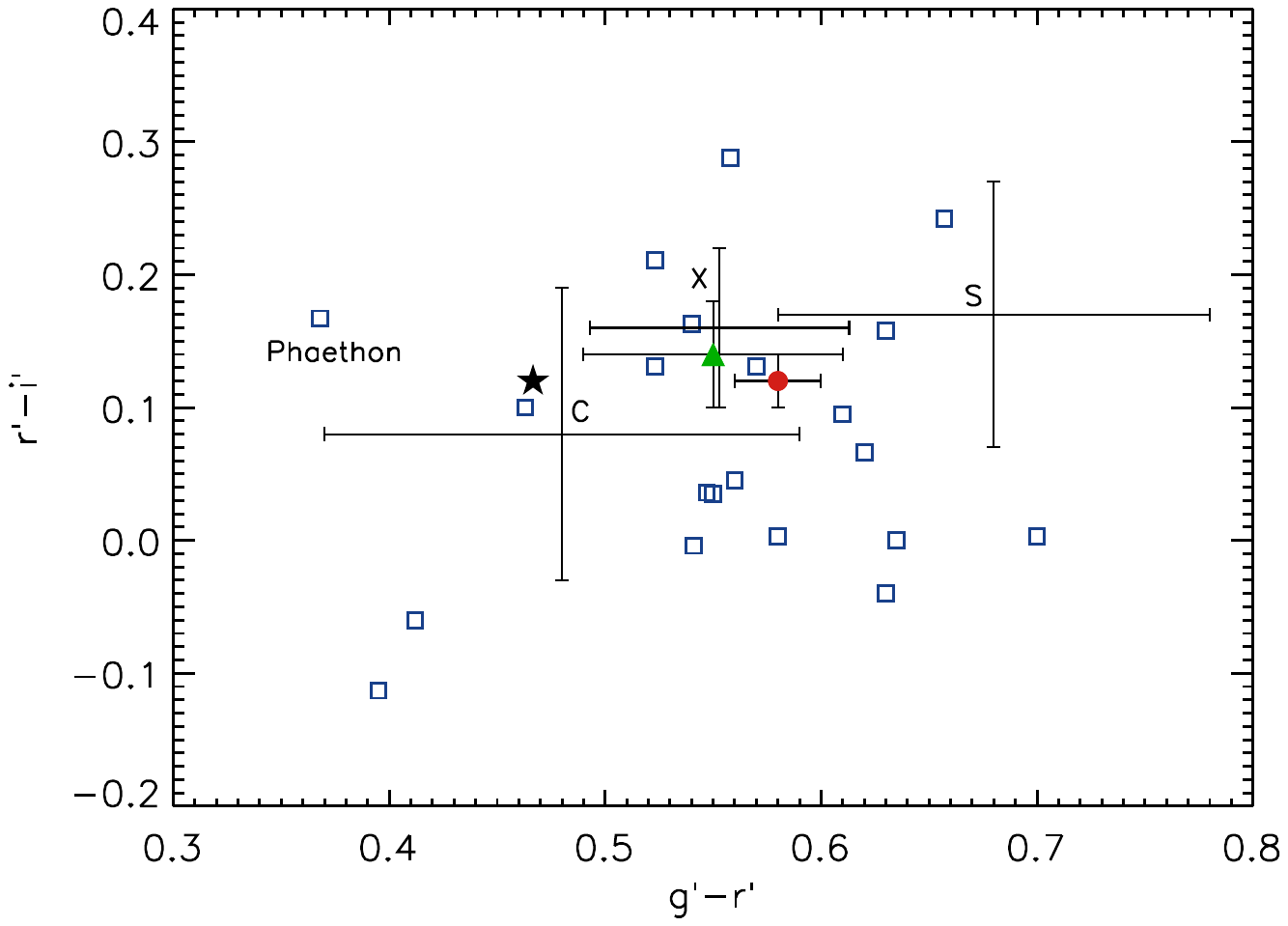}
\includegraphics[angle=180,totalheight=0.3\textheight]{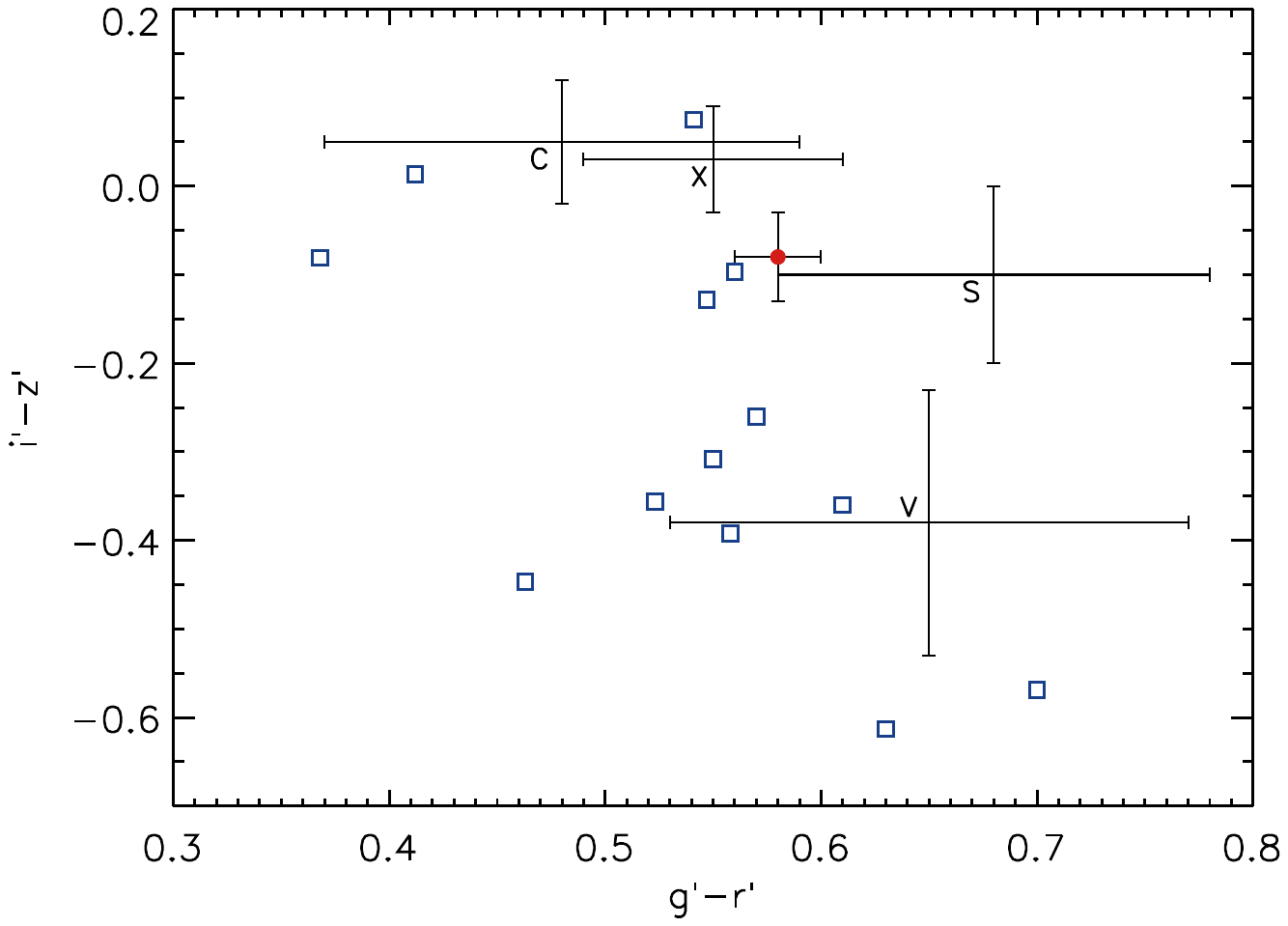}
%\vspace{-1.4in}
\caption{Sloan optical colors of low perihelion asteorids ($q < 0.15$
  au). 2021 PH27 is shown by the red filled circle and 2025 GN1 by the
  green filled triangle.  Other low perihelion asteroids are from
  Jewitt (2013) and Holt et al. (2022) and shown by blue squares
  (uncertainties have been removed for clarity, but are generally 0.05
  to 0.1 mags or better). The color of the Sun is a black star and the
  very blue Phaethon is labelled. The general range for the colors of
  the C-type, S-type, X-type and V-type asteroids are shown by a their
  first letter and range bars according to Paker et al. (2008) and
  Sergeyev \& Carry (2021). The z'-band allows the silicate feature to
  be probed as a turn over in the color, which is signified by a
  significant negative i'-z' color and the V-type asteroids according
  to Paker et al. (2008) and Sergeyev \& Carry (2021). Several of the
  low perihelion objects appear to have silicate features, though 2021
  PH27 does not.
\label{fig:colors} }
\end{figure}

Figure~\ref{fig:colors} shows the colors of other known low perihelion
asteroids from Jewitt (2013) and Holt et al. (2022). In general, the
colors of near-Sun asteroids seem to be as diverse as the population
of asteroids away from the Sun with colors ranging from blue to very
red (Jewitt 2013; Holt et al. 2022). 2021 PH27 and 2025 GN1 are on the
redder color side than most other low perihelion asteroids, but still
within the general colors of other objects.

As seen in Figure~\ref{fig:colors}, several of the low perihelion
asteroids have mafic or silicate features shown by their large
negative i'-z' color, though 2021 PH27 does not. (594913)
'Aylo'chaxnim (2020 AV2) is a larger than 1 km Vatira asteroid with an
orbit completely interior to Venus and found to have mafic spectral
features and classified as a Sa-type asteroid (Popescu et al. 2020).

3200 Phaethon is the well-known source of the Geminid meteors and has
a close passage to the Sun of 0.14 au at perihelion (Kasuga \& Masiero
2022; Henych et al. 2024). 3200 Phaethon is blue in color and
classified as a B-type asteroid and will be the target of the upcoming
Destiny+ spacecraft mission (Licandro 2007; Ozaki et al. 2022;
Beniyama et al. 2023). 3200 Phaethon shows activity when near the Sun,
suggesting the activity is from thermal stress fracturing that likely
occurs on some asteroids with perihelia inside Mercury's orbit (Jewitt
\& Li 2010; Li \& Jewitt 2013; Zhang et al. 2023). No obvious extended
emission or cometary features were seen as 2021 PH27 and 2025 GN1 were
point like sources with the same PSF as nearby field stars.

The surface regolith of asteroids close to the Sun might result from
thermal fatigue (Delbo 2014; MacLennan \& Granvik 2023). Lisse \&
Steckloff (2022) determined that a body subjected to the intense
thermal environment of the Sun, like 3200 Phaethon, would blue the
surface of such objects from the removal of Fe and other species that
tend to redden an object such as refractory organics. Lisse \&
Steckloff (2022) predict that objects with very small perihelia should
generally be blue. We find 2021 PH27 and 2025 GN1 do not show a blue
color as expected if they have lost significant refractory organics as
suggested by Lisse \& Steckloff (2022). Thus either this blueing
mechanism is not very prevalent or the creation of the 2021 PH27 and
2025 GN1 pair near the Sun was too recent for their colors to be
significantly modified by such a mechanism.

There appears to be a short-term light-curve for 2021 PH27 of about
0.1 magnitudes in the r, g and i bands over a period of about 0.77
hours on April 8, 2022. This suggests if 2021 PH27 is an elongated
object it would have a double-peaked period of at least 1.6 hours. On
April 9, 2022, the observations were only over about 0.54 hours and
show a similar apparent magnitude as the observations on April 9,
suggesting 2021 PH27 likely does not have a large amplitude light
curve, though a longer time-base of observations are needed to confirm
this trend. 2025 GN1 shows there might be a photometric amplitude
larger than about 0.2 mags over the 45 minutes of observations for
this object as it was consistently brighter at the end of the 45
minutes of observations.

\section{Orbital Dynamics}

Though 2025 GN1 does not have as precise orbital elements known as
2021 PH27 with only a few nights observational arc to date, we noticed
2025 GN1 does not just have a very similar semi-major axis,
eccentricity and inclination as 2021 PH27, but also a similar
longitude of ascending node and argument of perihelion. Thus the
longitude of perihelion, or place where 2025 GN1 reaches perihelion
near the Sun is less than one degree and thus almost in the exact same
location as 2021 PH27 (Table~\ref{tab:orbitalelements}). This is
somewhat unexpected as all but the semi-major axis of the orbital
parameters should rapidly change for the asteroids on a few to ten kyr
timescale due to strong perturbations by Mercury, Venus and Earth
(Sheppard et al. 2022).

In order to assess the dynamical orbits of 2021 PH27 and 2025 GN1, we
performed the exact same numerical orbit simulations as described in
Sheppard et al. (2022), but this time included the updated orbit of
2021 PH27 and the newly discovered 2025 GN1. For 2025 GN1, our new
astrometric measurements double the observational arc for this object
from the published discovery MPEC and the resulting improved orbit
determination used in the numerical simulations is shown in
Table~\ref{tab:orbitalelements}. For full details see Sheppard et
al. (2022), but briefly, in the numerical orbit simulation we used the
SWIFT RMVS4 integrator (Levison \& Duncan 1994) and included the
planets Mercury, Venus, Earth+Moon (barycenter), Mars (barycenter),
Jupiter (barycenter), Saturn (barycenter), Uranus (barycenter) and
Neptune (barycenter). The time step was 0.1 days and 1000 clones were
generated using a normal distribution within 3 sigma of the nominal
orbits of 2021 PH27 and 2025 GN1.

\begin{figure}
\epsscale{0.4}
\centerline{\includegraphics[angle=0,totalheight=0.4\textheight]{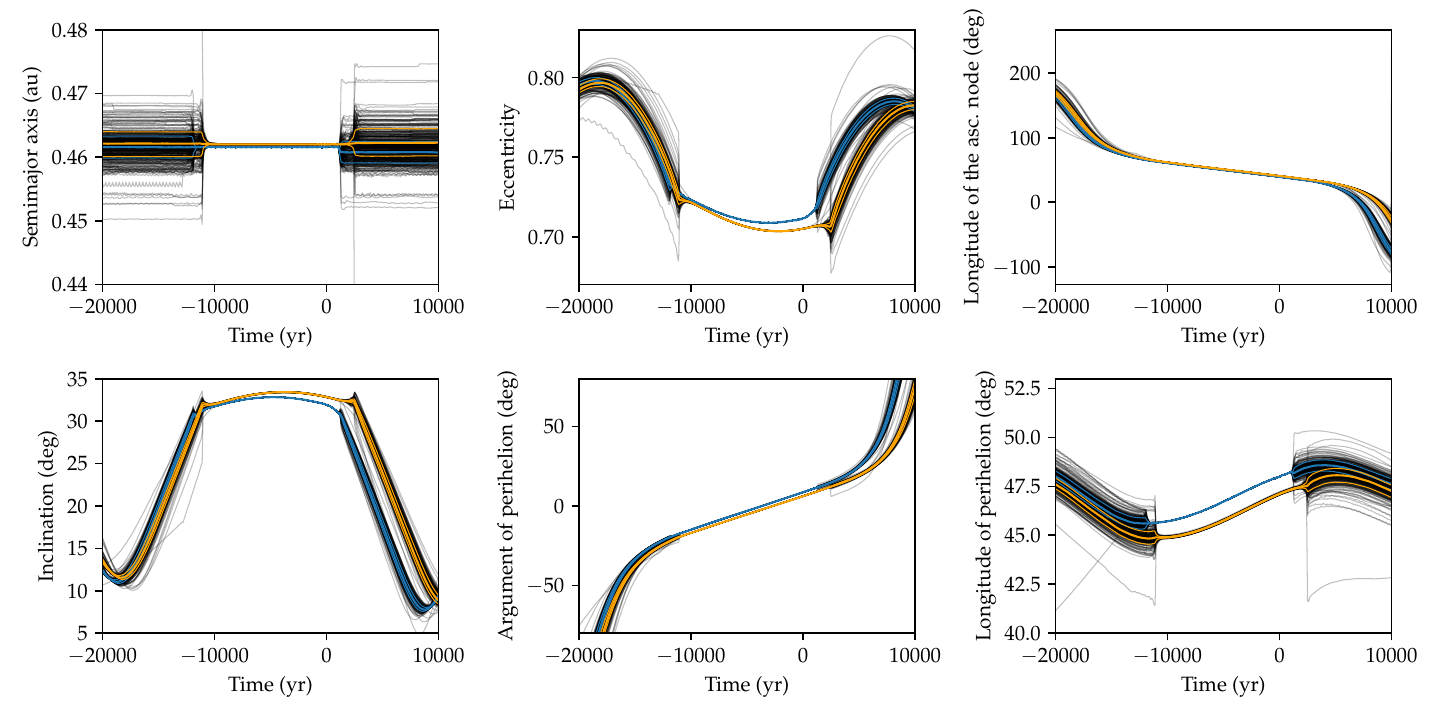}}
%\vspace{-1.4in}
\caption{Numerical orbital simulations of 2021 PH27 (blue) and 2025
  GN1 (yellow) and their clones over 30 kyrs starting 20 kyrs in the
  past. Bolded middle color lines show the median orbital parameters
  and the thin color lines the 1-sigma median value of each orbital
  element. 2021 PH27 experiences a very close encounter with Venus
  that is within 10s of Venus radii in about 1150 years (3175
  AD). 2025 GN1 experiences a very close encounter with Venus in about
  2155 years (4180 AD) with some clones even hitting Venus. The orbits
  of the two objects significantly diverge at these Venus
  encounters. Backwards orbital numerical integrations show 2021 PH27
  had another close encounter with Venus about -11,620 years ago while
  2025 GN1 may have encounter Venus around -11,000 years ago. These
  frequent encounters within the Hill sphere of Venus significantly
  alter the orbits of these objects and suggests that the 2021 PH27
  and 2025 GN1 pair was created around the last Venus encounter some
  11 kyrs ago or within the last several kyrs afterwards in order to
  still have such similar orbits that we see today.
\label{fig:dynamics} }
\end{figure}

Figure~\ref{fig:dynamics} shows the results of our numerical orbital
integrations. 2021 PH27 still has a very close encounter with Venus
within a few tens of Venus radii in about 1150 years (3175 AD). Though
2025 GN1's orbit is not as well determined, it appears it has a close
encounter within Venus' Hill sphere in about 2155 years (4180 AD), as
all clones are significantly altered in their orbits at this time and
some clones actually collided with the planet. These simulations show
the orbits of these two objects should significantly diverge by
several degrees after the Venus encounter by 2021 PH27 in about 1150
years time and again diverge in 2155 years time when 2025 GN1
encounters Venus.

Backward numerical orbital integrations are needed to allow us to
infer if 2021 PH27 and 2025 GN1 were likely together in the
past. Unfortunately, the orbital elements of 2025 GN1 are not yet at
the precision needed for an accurate backward numerical orbital
integration to assess the closeness of 2021 PH27 and 2025 GN1 in the
past and additional astrometry is needed at the next epoch to
determine how close 2021 PH27 and 2025 GN1 might have been to each
other in the past. But the orbit of 2021 PH27 is now very well known,
and backward numerical orbital integrations show it had a significant
encounter with Venus about 11620 years ago. This suggests the creation
of 2025 GN1 from 2021 PH27 likely occurred during this encounter or
within the last few to several kyrs in order for them to still have
such similar orbital elements that we see today as they appear to not
have been strongly perturbed by any of the planets since the pairs
formation. Though less certain, we find in the backward orbital
integrations 2025 GN1 had a very close encounter with Venus about
11000 years ago. Using Granvik et al. (2018), we find the favored
escaped route from the main asteroid belt for the parent object of
2021 PH27 and 2025 GN1 is likely the $\nu$ 6 resonance with a
probability of $91-97~\%$.

\section{Conclusions and Summary}

Both 2021 PH27 and 2025 GN1 were found to have similar Sloan optical
colors characteristic of an X-type asteroid, which overlaps and is
between the C-type and S-type asteroids in g', r' and i' color
space. z'-band imaging for 2021 PH27 does not show any mafic or
silicate features as found for some other low perihelion objects. 2025
GN1 is about 2.3 magnitudes fainter than 2021 PH27 in absolute
magnitude, making it around 400 km in diameter assuming a moderate
albedo and about a third of the size of 2021 PH27 if they both have
similar albedos.

The very similar orbits and colors of 2021 PH27 and 2025 GN1 suggest
they are fragments from a once larger parent asteroid or that 2021
PH27 has or is shedding material from the extreme stress environment
near the Sun. The fragmentation process may be ongoing, making
continual dust that might create an annual meteor shower at Venus. The
loss of mass from the main asteroid could be from thermal stress,
tidal stress, rotational spin-up or from an escaped satellite. 2021
PH27 does not show the very blue color as seen for asteroid 3200
Phaethon, which is the parent body of the Geminids meteor shower seen
on Earth. The strong planetary perturbations on 2021 PH27 and 2025 GN1
make their orbital parameters change significantly over several kyr
timescale, suggesting this pair seperated from each other only a few
to $\sim 10$ kyrs ago. 2025 GN1 will likely have a very close
encounter with Venus in about 2155 years (4180 AD).

An amplitude of about 0.1 mags was seen for 2021 PH27 over a 0.77 hour
time. The apparent magnitude of 2021 PH2 was similar on both nights of
observations of that object with no evidence of mass loss or activity
seen for 2021 PH27.  2025 GN1 showed a brightening of about 0.2 mag
short-term rotational variation over 0.75 hours of observations.

Many more low perihelion asteroids will likely be found with a NEO
surveyor type mission in the future (Masiero et al. 2024) or from
groundbased twilight surveys (Pokorny et al. 2020; Ye et al. 2020;
Sheppard et al. 2022; Schwamb et al. 2023; Bolin et al. 2025). Though
just a pair of objects now, more smaller members of a 2021 PH27
asteroid family may be expected to be found from these surveys,
depending on how this pair was created.

\section*{Acknowledgments}

This paper includes data gathered with the 6.5 meter Magellan
Telescopes located at Las Campanas Observatory, Chile. We thank Dave
Osip at Las Campanas Observatories for use of engineering time at the
Magellan telescope for some of the observations. PP acknowledges
support provided by NASA’s Planetary Science Division Research
Program, through ISFM work packages EIMM and Planetary Geodesy at NASA
Goddard Space Flight Center and NASA award number 80GSFC24M0006.

\clearpage

\newpage

\clearpage

%\begin{document}

\begin{center}
\begin{deluxetable}{llcccc}
%\small
\tablenum{1}
\tablewidth{6 in}
\tablecaption{Geometrical Circumstances of the Observations \label{tab:GeoPhase}}
\tablecolumns{6}
\tablehead{
\colhead{Name} & \colhead{UT Date} & \colhead{$R$}  & \colhead{$\Delta$} & \colhead{$\alpha$} & \colhead{Elong} \\ \colhead{} &\colhead{} &\colhead{(AU)} &\colhead{(AU)} & \colhead{(deg)} & \colhead{(deg)} }  
\startdata
2021 PH27                        & 2022 Apr $08.383-.418$  & 0.767 & 0.593 & 93.9 & 49.8 \\
2021 PH27                        & 2022 Apr $09.389-.417$  & 0.763 & 0.591 & 94.6 & 49.4 \\
2025 GN1                         & 2025 Apr $14.387-.417$  & 0.750 & 0.554 & 99.5 & 47.5 \\
\enddata
\tablenotetext{}{Quantities are the heliocentric distance ($R$), geocentric distance ($\Delta$), phase angle ($\alpha$) and Elongation from the Sun (Elong). UT Date shows the year, month, day and time span of the observations.} 
\end{deluxetable}
\end{center}

%\end{document}             % End of document.
  %\label{tab:GeoPhase} Geometric Circumstance Observations

%\newpage

%\input{SheppardTableB.tex}  %\label{tab:mags} apparent mags in sloan filters.

\newpage

%\documentstyle [aj_pt4]{article}    % Specifies the document style.

%\begin{document}

\begin{center}
\begin{deluxetable}{lcccccc}
%\small
\tablenum{2}
\tablewidth{6.0 in}
\tablecaption{Average Sloan Colors \label{tab:Colors}}
\tablecolumns{7}
\tablehead{
\colhead{Type} & \colhead{g'-r'} & \colhead{r'-i'} & \colhead{i'-z'} & \colhead{g'-i'} \\ \colhead{} & \colhead{(mag)} & \colhead{(mag)} & \colhead{(mag)} & \colhead{}}  
\startdata
2021 PH27  &   $0.58 \pm 0.02$    &   $0.12 \pm 0.02$    &   $-0.08 \pm 0.04$ & $0.70 \pm 0.03$ \\
2025 GN1   &   $0.55 \pm 0.06$    &   $0.14 \pm 0.04$    &   $-$              & $0.69 \pm 0.07$ \\
\enddata
\tablenotetext{}{}
\end{deluxetable}
\end{center}

%\end{document}             % End of document.

  %\label{tab:Colors} sloan colors.

\newpage

%\documentstyle [aj_pt4]{article}    % Specifies the document style.

%\begin{document}

\begin{center}
\begin{deluxetable}{lccccccccc}
%\small
\tablenum{3}
\tablewidth{6.5 in}
\tablecaption{Orbital Elements \label{tab:orbitalelements}}
\tablecolumns{10}
\tablehead{
\colhead{Name} & \colhead{$q$}  &  \colhead{$a$} & \colhead{$e$}  & \colhead{$i$} & \colhead{$\Omega$} & \colhead{$\omega$} & \colhead{$\bar{\omega}$}   & \colhead{Dia}  & \colhead{$H_{r}$} \\ \colhead{} & \colhead{(AU)} & \colhead{(AU)}  & \colhead{} &\colhead{(deg)} & \colhead{(deg)} &\colhead{(deg)} & \colhead{(deg)} & \colhead{(m)}  & \colhead{(mag)} }  
\startdata
2021 PH27  & 0.13319   & 0.46176 & 0.7116 & 31.9411 &  39.397 & 8.5796 & 47.976 &  1200 & 17.6 \\
2025 GN1   & 0.136     & 0.462   & 0.705  & 32.8    &  41.0   & 6.1    & 47.1   &  400  & 19.9 \\
\enddata
\tablenotetext{}{Orbital parameters of 2021 PH27 are from the Minor Planet Center as of April 2025. Quantities are the perihelion ($q$), semi-major axis ($a$), eccentricity ($e$), inclination ($i$), longitude of the ascending node ($\Omega$), argument of perihelion ($\omega$), and longitude of perihelion ($\bar{\omega}$). Diameter (Dia) estimates assume a moderate albedo of 0.15.  Uncertainties are the number of significant digits.}
\end{deluxetable}
\end{center}

% magnitude, size

%\end{document}             % End of document.

  %\label{tab:orbitalelements} orbital elements

\newpage

\end{document}